# Halogen-Bond Driven Self-Assembly of Perfluorocarbon Monolayers

Antonio Abate,[1, 2] Raphael Dehmel,[1] Alessandro Sepe,[1, 2] Ngoc Linh Nguyen,[3] Bart Roose,[2] Nicola Marzari,[3] Jun Ki Hong,[4] James M. Hook,[5] Ullrich Steiner,[2] Chiara Neto* [4]

[1] Department of Physics, Cavendish Laboratory, JJ Thomson Avenue, Cambridge, CB3 0HE, United Kingdom.

[2] Adolphe Merkle Institute, University of Fribourg, Chemin des Verdiers 4, CH-1700 Fribourg, Switzerland

[3] Theory and Simulations of Materials (THEOS), and National Centre for Computational Design and Discovery of Novel Materials (MARVEL), École Polytechnique Fédérale de Lausanne, 1015 Lausanne, Switzerland

[4] School of Chemistry and Australian Institute for Nanoscale Science and Technology, The University of Sydney, NSW 2006 Australia

[5] Mark Wainwright Analytical Centre, University of New South Wales, Sydney, NSW, 2052, Australia

* Corresponding author: chiara.neto@sydney.edu.au

**Abstract**

The self-assembly of a single layer of organic molecules on a substrate is a powerful strategy to modify surfaces and interfacial properties. The detailed interplay of molecule-to-substrate and molecule-molecule interactions are crucial for the preparation of stable and uniform monomolecular coatings. Thiolates, silanes, phosphonates and carboxylates are widely used head-groups to link organic molecules to specific surfaces study we show that self-assembly of stable and highly compact monolayers of perfluorocarbons. Remarkably, the lowest ever reported surface energy of 2.6 mJ $m^{-2}$ was measured for a perfluorododecyl iodide monolayer on a silicon nitride substrate. As a convenient, flexible and simple method, the self-assembly of halogen-bond driven perfluorocarbon monolayers is compatible with several applications, ranging from biosensing to electronics and microfluidics. Compared to other methods used to functionalise surfaces and interfaces, our procedure offers the unique advantage to work with extremely inert perfluorinated solvents. We demonstrate that surfaces commonly unstable in contact with many common organic solvents, such as organic-inorganic perovskites, can be functionalized via halogen bonding.

The self-assembly of organic monolayers (SAMs) is an established strategy to modify surfaces and interfacial properties.[1] SAMs can easily form through the spontaneous, rapid adsorption of small molecules from solution or the gas phase.[2] The interplay between molecule-to-molecule and molecule-to-substrate interactions controls the packing of the molecules on the surface and two-dimensional, ordered domains can be prepared by self-assembly for a large variety of small molecules on solid surfaces. SAM formation typically relies on a molecular head-group with a specific, strong affinity for the substrate, which leads to stable and compact monomolecular layers. The most common SAMs exploit the interaction of thiols with gold and other noble metals, silanes and phosphonates with silica or carboxylates with alumina and titania.[3-5] SAMs are often used to modify wettability, corrosion resistance, adhesion, friction, conductivity and bio-compatibility, and to elicit specific biological responses. As a fundamental field of research, they have prompted a very wide range of basic and applied studies over the past twenty-five years, ranging from surface-force measurements to biosensing and molecular electronics. Lastly, SAMs can also be easily patterned through soft-lithography approaches, which have been used as a platform for numerous applications, including the formation of arrays of single cells and open microfluidics. New strategies of SAM formation are therefore likely to lead to further advances in nanoscience and nanotechnology.

Here, we demonstrate the use of non-covalent halogen bonding as the driving molecule-to-substrate interaction for the formation of perfluorocarbon-based SAMs. Several experimental and modelling approaches provide evidence for the stable and reproducible formation of monomolecular perfluorocarbon layers on substrates via such halogen bonding. In addition, these perfluorocarbon-based SAMs give rise to high fluorine densities, which result in extremely low surface energies and large water-contact angles. Different from other methods used to functionalise surfaces and interfaces, our procedure offers the unique advantage to work with low molecular weight perfluorocarbons, which are extremely inert and immiscible with many common hydrophilic and oleophilic solvents. Materials that are unstable in contact with common solvents, and thus difficult to functionalize with traditional methods, such as organic-inorganic perovskites, are compatible with our procedure.

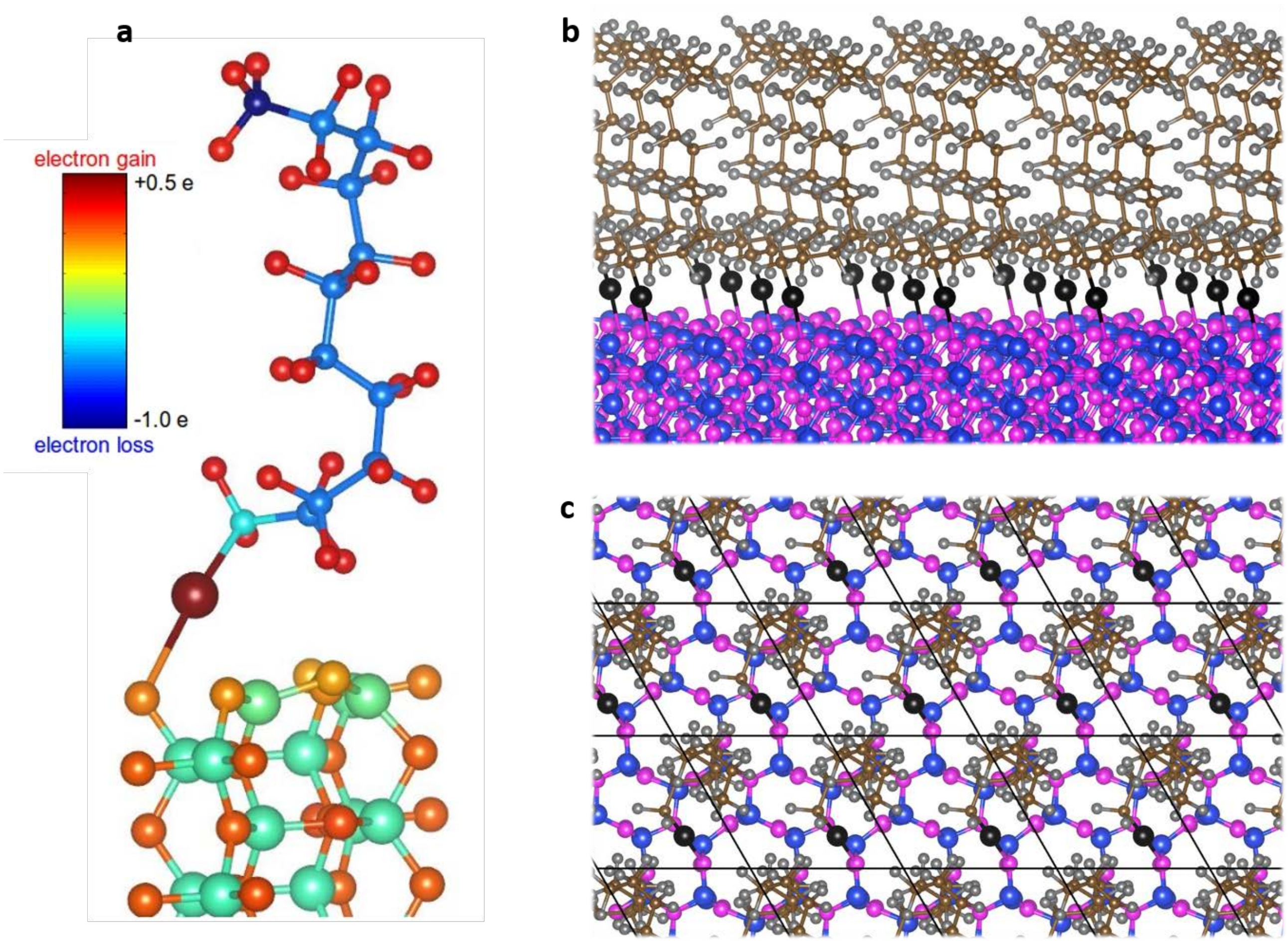

**Figure 1**. a) Löwdin atomic charge analysis of perfluorododecyl iodide (I-PFC12) on a (001) silicon nitride surface as calculated by a density functional theory (see Experimental Section for details). The terminal I atom the I-PFC12 is linked to the N atom on the substrate surface. The colour of the atoms indicates the electron density redistribution upon the interaction of I-PFC12 with the silica nitride substrate: red (blue) corresponds to electron gain (loss). b) Side view and (c) top view of I-PFC12 on a silicon nitride substrate with the I-PFC12 molecules on the surface as calculated by density functional theory. Element colour codes in b and c: I, black; C, brown; F, white; N, violet; Si, blue.

The term 'halogen bond' defines an attractive non-covalent interaction between an electrophilic region associated with a halogen atom in a molecular entity and a nucleophilic region in another, or the same, molecular entity.[6-9] Here, we employ the electrophilic terminal iodine in the perfluorododecyl iodide molecule (I-PFC12) to enable the donation of electron density from the nucleophilic nitrogen atoms of an inorganic silicon nitride ($SiN_x$) substrate. Bertani *et al.* described halogen bonding for the perfluoro-functionalization of nitrogen-containing organic surfaces,[10] but no similar reports for inorganic surfaces are yet to appear.

Density functional theory (DFT) was used to explore potential molecule-to-substrate interactions, using simplified model assumptions. In particular, predefined substrate absorption sites and molecular geometries were used to simplify the calculation. This DFT model is useful to elucidate intermolecular binding mechanisms in the SAM in terms of electron-density distributions, but it does not allow conclusions to be drawn about the molecule-to-substrate binding mechanism, the actual structure of molecules on real surfaces, the density of the monolayer, nor its long-term stability. This evidence is deduced from spectroscopic techniques (vide infra).

Figure 1a shows a Löwdin atomic charge analysis of the molecule-substrate system.[11] The electron density redistribution of I-PFC12 on the $SiN_x$ indicates that approximately 0.5 electrons were transferred from the nucleophilic nitrogen surface atom to the electrophilic iodine atom, with respect to its neutral valence charge value (7 electrons). Among the other halides (such as bromine and chlorine), we focused on iodine as the strongest halogen bonding donor.[12] The calculation yields values of 2.205 and 2.791 Å for the distances between carbon-iodine and iodine-nitrogen respectively. These are bond lengths that are typical of perfluorocarbons involved in a halogen bonding interaction with a nitrogen lone pair on another molecular entity.[12] Similar to earlier reports, the interaction between the I-PFC12 and the $SiN_x$ is primarily electrostatic in nature, although second order contributions such as polarization, dispersion and charge transfer are present. Figure 1b shows the side views of I-PFC12 molecules assembled on $SiN_x$. The results of the calculation show that non-covalent halogen bonding is the dominant interaction between I-PFC12 and (001) silicon nitride, with a binding energy of 128 kJ/mol. Similar to previous reports, our calculations indicatecalculation indicates that the occurrence of gauche conformations within I-PFC12 (as shown in Figure 1a) results in a higher molecule-to-substrate binding energy compared to a fully trans conformation (see SI).[13-14] Note that this theoretical value is higher than the strength commonly reported for thiols on gold (84 kJ/mol)[15], which is one of the most commonly used molecule-to-substrate interactions in the preparation of SAMs. In Figure 1c, the top view of I-PFC12 assembled on silicon nitride illustrates the possibility of a highly-ordered monolayer, with a molecule orientation perpendicular to the substrate.

I-PFC12 was also modelled on a silicon substrate,[16] exposing the formation of covalent iodine-silicon bonds, caused by the hybridization between the valence electrons of the iodine and silicon atoms (see SI). In comparison to $SiN_x$, halogen bonding of I-PFC12 onto silicon oxide is more ionic (see SI). DFT modelling also predicts a less-ordered monolayer for I-PFC12 that is covalently bound to the substrate. A higher order may allow a more densely packed monolayer.[17] Indeed, the calculation yields values of 7.57 Å and 10.86 Å for molecule-to-molecule distance in the halogen and covalently bonded SAM, respectively. Therefore, non-covalent halogen bonding may offer an intrinsic advantage when preparing densely packed SAMs. The study of I-PFC12 on silicon is still under investigation and it will be published elsewhere.

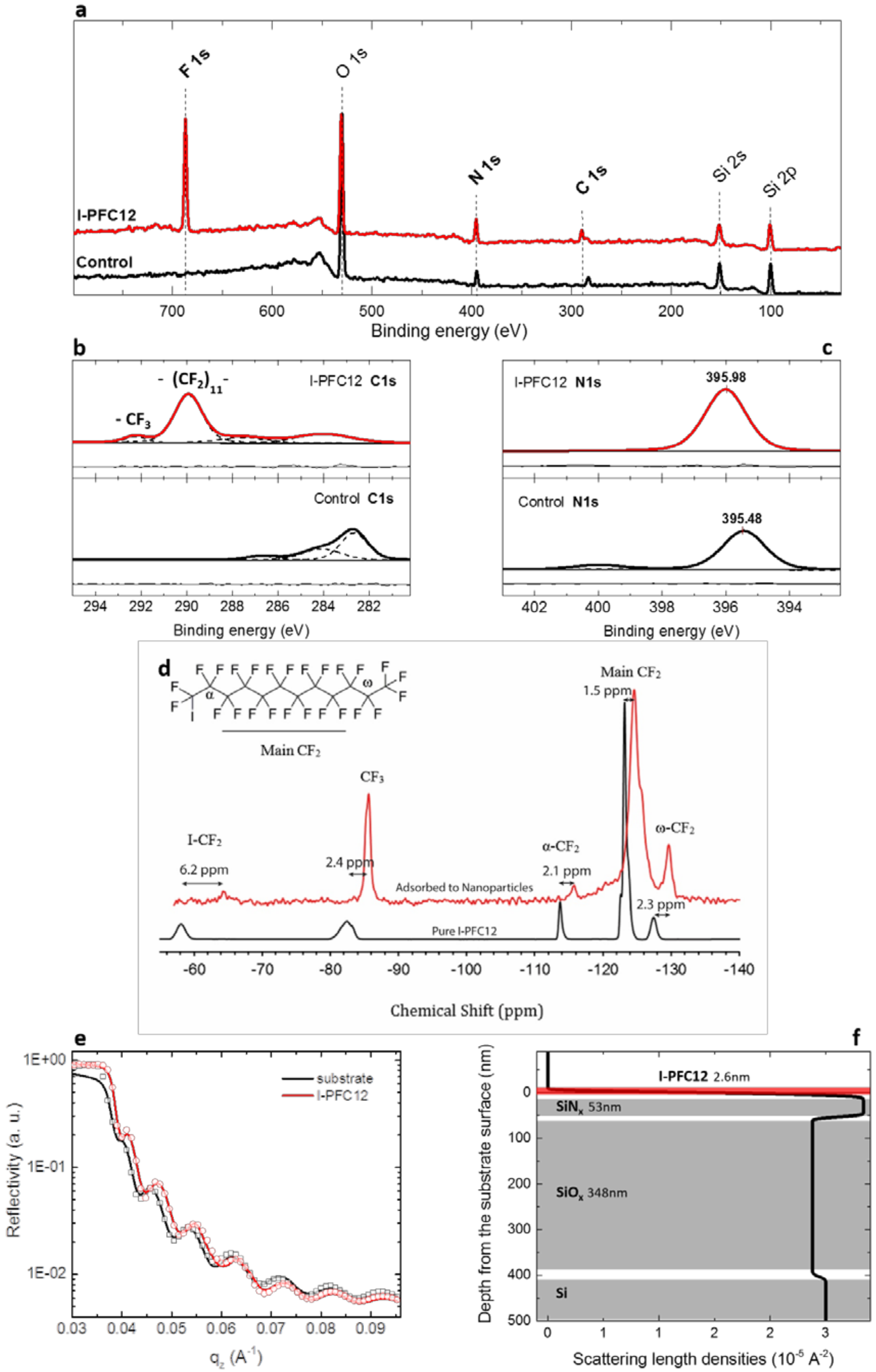

**Figure 2**. a) XPS spectra of silicon nitride substrates that were immersed in a solution of tetradecafluorohexane with and without (control) perfluorododecyl iodide (I-PFC12). The high-resolution spectra in (b) and (c) have the same intensity scale. The dashed lines are fits extracted from a Gaussian peak deconvolution. The residuals of these fits are plotted below the spectra. d) $^{19}F$ MAS NMR spectra of pure I-PFC12 (black) and I-PFC12 self-assembled monolayer adsorbed to $SiN_x$ nanoparticles (red). The spectra were acquired at $B_0 = 16.44$ T at room temperature with a MAS spinning frequency of 60 kHz and 2000 acquisitions. Inset: Molecular structure of IPFC-12 with labeled $CF_x$ fragments. Trifluoroacetamide was used as an external reference (see SI). e) Synchrotron X-ray reflected intensity as a function of the photon momentum change $q_z$ upon reflection, for bare and I-PFC12-covered silicon nitride substrates. The solid lines show fits to the data corresponding to the multilayer model with the scattering length density shown in (f). The substrate layer sequence and their thicknesses are schematically indicated, showing a stack of Si, $SiO_x$, $SiN_x$ and I-PFC12.

Halogen bonded perfluorocarbon SAMs were manufactured by immersing $SiN_x$ substrates in an I-PFC12 solution in tetradecafluorohexane. As soon as they were removed from the I-PFC12 solution, the substrates were first rinsed and then immersed in fresh tetradecafluorohexane to remove the excess of I-PFC12. Figure 2a shows the X-ray photoelectron spectroscopy (XPS) analysis of $SiN_x$ substrates with and without I-PFC12 treatment, providing the elemental composition and bonding states of the surfaces. The predominant peaks of the surface treated with pure tetradecafluorohexane (black line in Figure 2a, control) correspond to O, N, C and Si. The N and Si peaks are expected for $SiN_x$; the Si 2p peak centred at 101.7 eV is a clear signature of Si–N bonds within the $SiN_x$ lattice.[18] The presence of C and O peaks presumably stems from residual adventitious surface contamination and oxidation (see XPS depth profile in the SI for more detail discussion about the presence of the oxygen peak). The spectrum of the I-PFC12 treated substrates (red line in Figure 2a, I-PFC12) shows an additional sharp peak at 687 eV, which is attributed to the presence of F-C bonds.[19] The integration of this peak yields an estimated value of about 20% for the fluorine content of the total surface elemental composition (see SI). The control substrate does not show any significant signal in this spectral region. This indicates that the fluorine signal does not originate from residues of tetradecafluorohexane (in which both samples were rinsed), and must therefore be attributed to the presence of I-PFC12, which was not removed by extended immersion of the sample in tetradecafluorohexane. Note that the I 3d signals of I-PFC12 around 620 and 632 eV are below the detection limit of the XPS, which is typically 0.5 atomic % of the 10 nm deep surface layer.[20] Figure 2b shows high resolution XPS spectra of the C1s signal. Substrates with and without I-PFC12 show traces in the spectral region between 281 and 288 eV. These traces are too broad to be assigned to specific bond states, particularly since they are likely to arise from adventitious contamination caused by the substrate or the atmosphere. The I-PFC12 treated substrate however showed two additional peaks at 290 and 292 eV that were assigned to $CF_2$ and $CF_3$ within perfluoroalkyl chains.[19] The ratio of these the two peaks was 10.3 (see SI), which is in good agreement with the $CF_2$ to $CF_3$ ratio in I-PFC12. This is a further indication of I-PFC12 adsorption onto the $SiN_x$ surface. Figure 2c shows the high-resolution XPS spectra of the N1s signal. The Gauss-fitted main peak position shows an increase of the binding energy by 0.5 eV for the substrate treated with I-PFC12. This agrees with the change in electronic density caused by the charge transfer of the lone electron pair from surface nitrogen atoms to the adsorbed I-PFC12 molecules through halogen bonding (see Figure 1).[21] Note that the N1s peak of the I-PFC12 substrate combines the signal from the surface and the in the top 10 nm of the $SiN_x$ substrate, which masks the larger energy shift of earlier reports. The control substrate also exhibited a small broad peak around 400 eV, which was not detectable in the I-PFC12 samples. Peaks around 400 eV were attributed to nitrogen bonded to $sp^3$ carbon (N–$sp^3$ C) and to nitrogen linked to $sp^2$ carbon (N–$sp^2$ C), which may arise from the binding of the surface carbon contamination to nitrogen, in agreement with the carbon signals in Figure 2b.[22] The I-PFC12 treatment therefore seems to protect the substrate surface by displacing traces of contaminating carbon compounds. It is possible that the I-PFC12 molecules halogen bond also to the surface oxygen, which will be investigated elsewhere. However, silicon nitride substrates with a lower oxygen content (non-plasma treated, see SI) also revealed monolayer formation.

$^{19}F$ solid-state nuclear magnetic resonance spectroscopy (ssNMR) was employed to gather further highly sensitive technique[23-27] to provide evidence of halogen-bonding of I-PFC12 to $SiN_x$.[23] Figure 2d shows the ssNMR spectra of pure I-PFC12 and a I-PFC12 self-assembled monolayer adsorbed to $SiN_x$ nanoparticles. The chemical shifts were assigned guided by previous according to the literature.[10, 23, 27] The simultaneous occurrence of the following features provides strong evidence for halogen bonding: i) the large upfield shift of the $CF_2$ groups bound to the iodine atom (I-$CF_2$, $\Delta\delta$ = -6.2 ppm) and the 2.1 ppm shift of the α-$CF_2$; ii) the upfield chemical shifts of the terminal $CF_3$ group ($\Delta\delta$ = 2.4 ppm); iii) the line width narrowing of the terminal $CF_3$ peak, which suggests an increased mobility at the monolayer/air interface with respect to the other fluorine groups; iv) the line width broadening of the main $CF_2$ backbone which suggests a reduced chain mobility due to adsorption onto the solid surface, and packing together.[24-27]

The XPS and the ssNMR analysis demonstrate that I-PFC12 molecules are adsorbed to $SiN_x$ through halogen bonding, forming a uniform organic layer (see also the atomic force microscopy images in the SI). Further evidence on the nature of the interaction came from solution state $^{19}F$ NMR analysis of washings of the functionalised silicon nitride microparticles, which returned the IPFC12 unchanged and demonstrated that the interaction is non-covalent and occurs only for molecules that contain a terminal iodine atom. X-ray reflectometry was employed to show that I-PFC12 adsorption results in the formation of monomolecular layers of reproducible thickness. The intensity of the reflected x-rays is sensitive to the electron density of different surface layers, allowing the extraction of the scattering length density profile perpendicular to the substrate with sub-nanometre accuracy.[5] Figure 2e shows the synchrotron x-ray reflection (XRR) intensity as a function of the momentum change of the photons upon reflection from $SiN_x$ substrates with and without I-PFC12 treatment. These data were analysed by comparing them to a simulated multilayer model that incorporates several variable parameters including the film thicknesses, layer compositions, and interfacial roughness. The XRR results were interpreted using a model calculation employing the Parratt formalism. The scattering length density (SLD) of Si, as well as the SLD and thickness of the $SiO_x$ and $SiN_x$ layer were fitted by the algorithm: an SLD value of 2.3 × 10-5 $Å^{-2}$ and a thickness of 348 nm were obtained for the $SiO_x$ layer, while an SLD of 2.8 × 10-5 $Å^{-2}$ and a thickness of 53 nm were obtained for the $SiN_x$ (Figure 2f). The substrates used were silicon wafers with an approximately 350 nm thick silicon oxide layer and a 50 nm thick $SiN_x$ surface layer. For the I-PFC12 treated substrates a good fit could only be obtained when including an additional 26 ± 4 Å thick top layer into the model with a scattering length density of $1.7 \times 10^{-5}$ $Å^{-2}$. This scattering density is typical for densely packed perfluorinated materials.[28] Based on the rigid rod-like structure of perfluorinated molecules[29] and the measured monolayer thickness, we conclude that the perfluoroalkyl chains stand perpendicularly on the substrate, forming a single dense molecular layer. This is in good agreement with the DFT picture proposed in Figure 1.

**Table 1**. I-PFC12 thickness on silicon nitride extracted from the synchrotron x-ray reflectometry data after immersion in three solvents for 14 hours.

| Category | Solvent | SAM thickness (Å) |
|---|---|---|
| Fluorinated | Tetradecafluorohexane | 26 ± 4 |
| Protic | Ethanol | 9 ± 4 |
| Aprotic | Toluene | 24 ± 4 |

SAMs are useful only if they are robust when exposed to various liquids.[30] To assess the stability of I-PFC12 SAMs, they were immersed overnight in three different solvents, followed by the measurement of the remaining monolayer thickness by X-ray reflectometry. Table 1 lists the thickness of the top layer after immersion in tetradecafluorohexane (fluorinated), ethanol (protic), and toluene (aprotic). The SAM thickness remained almost unaffected upon exposure to the fluorinated and aprotic solvents, but a thickness change was observed in protic solvents, such as ethanol, which washes off the monolayer but only after several hours of soaking. This observation can be explained by the competition of the hydrogen bonding that these solvents can establish with the nitrogen of the substrate, which could replace halogen bonded SAMs. The resistance against the protic solvent however indicates the good integrity of the I-PFC12 SAM, substantially delaying the penetration of ethanol and thereby enhancing its lifetime against a competing solvent. Complete studies about the competition between halogen and hydrogen bonding have been reported elsewhere.[31]

In absence of protic solvents, we found that I-PFC12 SAMs were perfectly stable over several weeks. Prolonged exposure to light may however promote substrate-mediated photochemical reactions which displace the iodide from the molecule, but do not lead to the detachment of the I-PFC12 from the substrate (see SI).[32-33] Interestingly, we found that I-PFC12 SAMs on micrometre size silicon nitride particles were stable to light exposure, as in this system most of the surface is buried and not exposed to light (see SI). A 10 mM IPFC-12 solution in tetradecafluorohexane exposed to light for one month revealed no significant changes to the chemical shifts of IPFC-12 (see SI). Hence, the displacement of iodide is probably a substrate-mediated process.

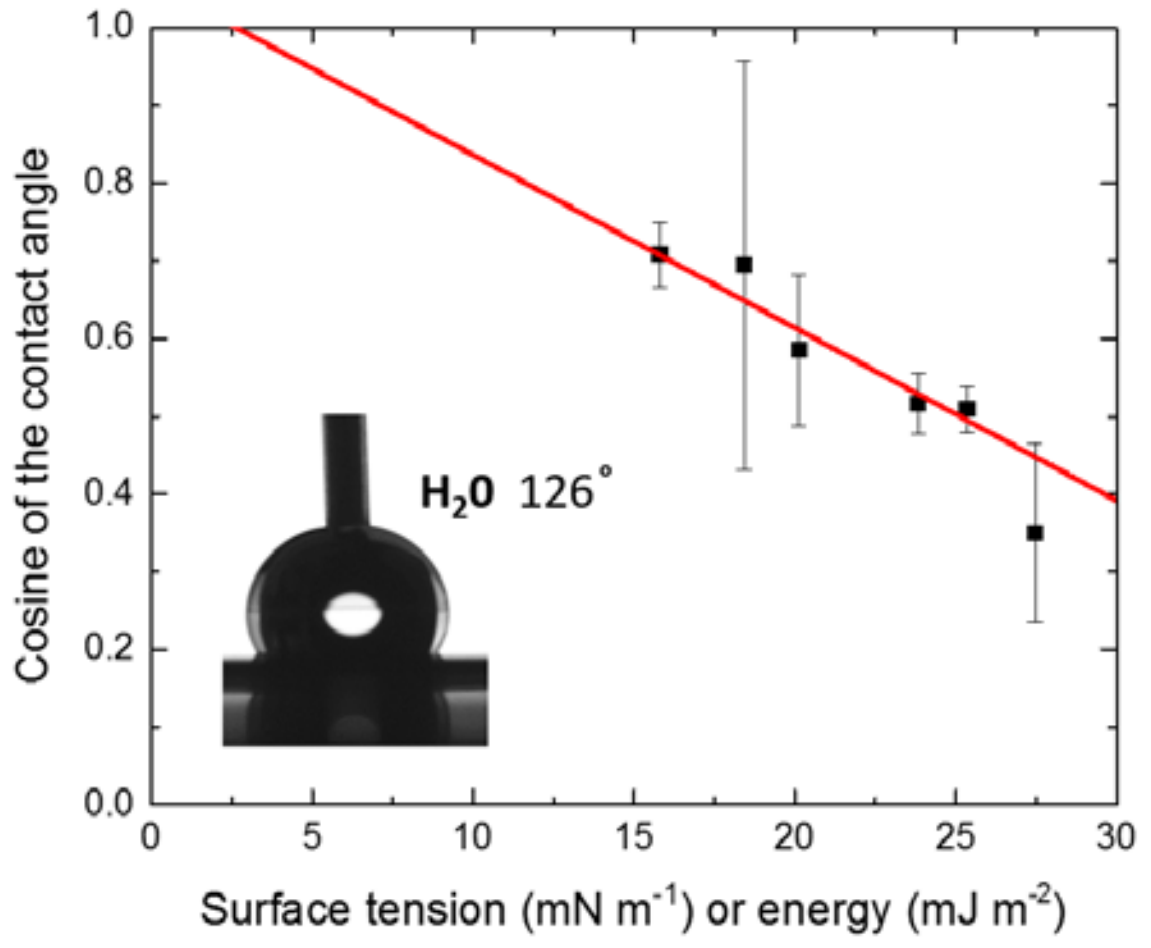

**Figure 3**. Zisman plot of I-PFC12 treated silicon nitride substrates. Each point is the average of 7 independent measurements collected with a known probing liquid. In order of increasing surface tension, the probing liquids were: pentane, hexane, heptane, decane, dodecane, hexadecane. A critical surface energy of 2.6 mJ $m^{-2}$ was extracted by linear extrapolation of the cosine of the contact angle to 1 (line). Inset, advancing contact angle of water on an I-PFC12 treated substrate.

Fluorocarbons are well known for their ability to repel both organic and water-based liquids.[34] For instance, fluorinated polymers have been widely used as coatings to manufacture low-stick surfaces with critical surface energies as low as 14 mJ $m^{-2}$. Poly(tetrafluoroethylene), also known under the commercial name Teflon®, for example, has a surface energy of approx. 20 mJ $m^{-2}$.[35] Significantly lower energies have been achieved preparing highly uniform surfaces with a large density of $CF_3$ groups.[36-38] The remarkably low surface energy of 6 mJ $m^{-2}$ has been reported for SAMs of perfluorolauric acid on platinum.[39]

Here, the critical surface energy of I-PFC12 SAMs was determined by the Zisman method.[40] In this method, drops of a series of alkanes were placed onto the substrate, and the advancing contact angle was determined. The surface energy of the substrate was then determined by extrapolating to a contact angle of zero. This is shown in Figure 3, where the cosine of the contact angle is plotted versus the surface energy for the homologous alkane series (pentane, hexane, heptane, decane, dodecane, hexadecane). The linear extrapolation of these data (averaged over seven independent experiments) results in the critical surface energy of the I-PFC12 treated $SiN_x$ surface of 2.6 mJ $m^{-2}$. To the best of our knowledge, this is the lowest ever reported surface energy of a solid surface. This exceptionally low value indicates an extremely large surface density of the $CF_3$ groups, which can be achieved only in highly uniform and densely packed fluorinated SAMs.[41]

To verify the Zisman extrapolation in Figure 3, a linear extrapolation was used to predict the contact angle of water on the I-PFC12 SAM. Taking the surface tension of deionised water as 72 mJ $m^{-2}$,[42] a contact angle of 123° is predicted. The inset in Figure 3 shows an image of a deionised water drop on I-PFC12 covered $SiN_x$, with an advancing contact angle of 126°. This value is in good agreement with our estimation, and it corroborates the extremely low surface energy extrapolated from the Zisman plot. Note that the water contact angle on I-PFC12 SAMs is significantly higher compared to other fluorinated SAMs, which have been prepared by adsorption of fluorinated alkanethiols onto gold surfaces (115°).[43] Halogen bonding therefore clearly enables more densely packed self-assembly than other methods, which are commonly used to manufacture SAMs.

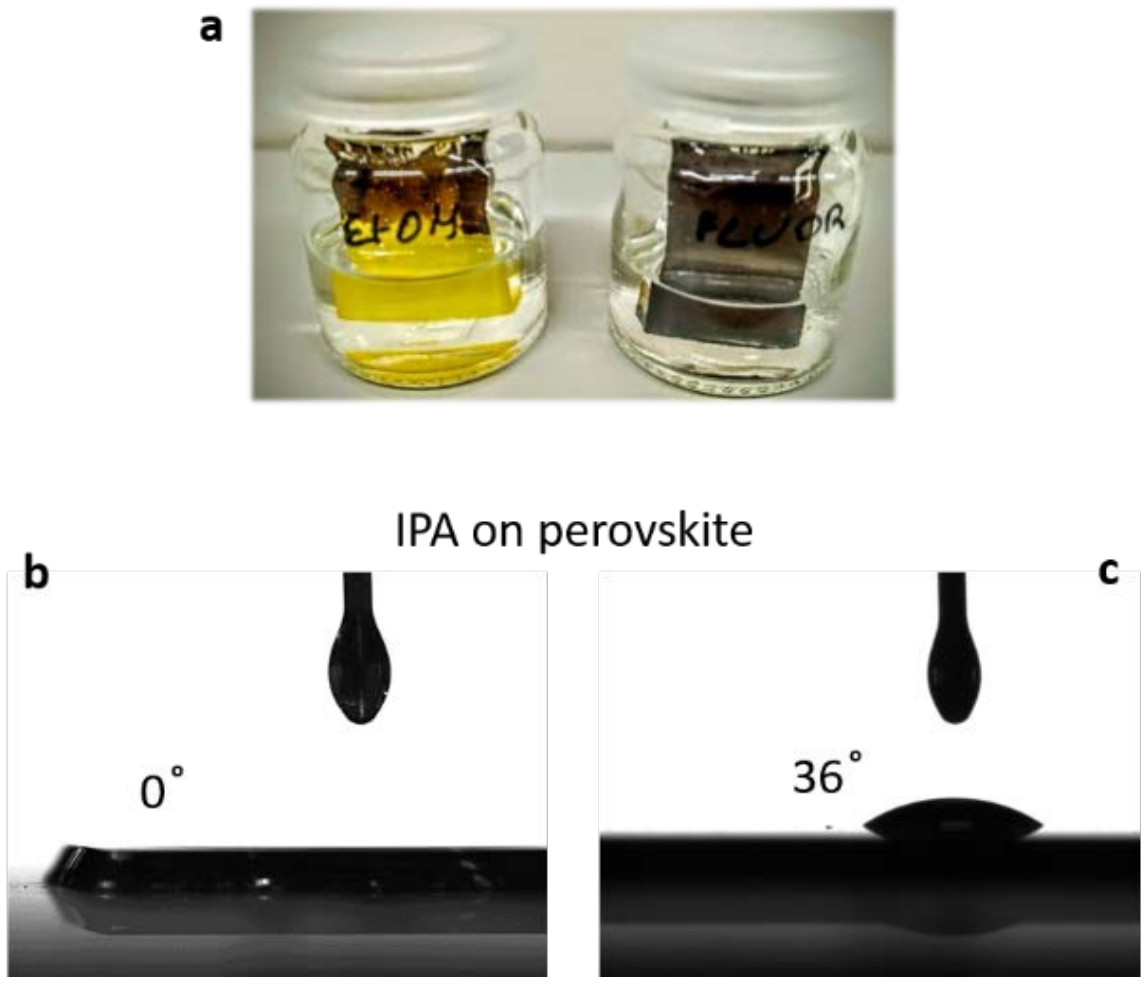

**Figure 4**. a) Uniform dark film formed by the organic-inorganic methylammonium lead iodide perovskite. The substrates were then immersed overnight at room temperature in pure ethanol (EtOH, left) and tetradecafluorohexane (FLUOR, right). The yellow colour is indicative of degradation of the perovskite film. Equilibrium contact angle of isopropanol (IPA) on (b) an untreated perovskite substrate, and (c) an I-PFC12 treated perovskite. Both the films were washed with tetradecafluorohexane before the measurement.

Even though the focus of this work is on the halogen-bond driven self-assembly from a fundamental point of view, this special type of monolayer is particularly useful for functionalizing surfaces that are unstable in contact with common organic solvents. Perfluorinated solvents, such as the tetradecafluorohexane used here, are well known for being extremely inert and immiscible with many common hydrophilic and oleophilic solvents.[44] Materials that are unstable in contact with common solvents and are thus difficult to functionalize with traditional methods are compatible with the procedure described above. Figure 4a provides clear evidence of this, showing methylammonium lead iodide perovskite films deposited on glass.[45] This and similar organic-inorganic perovskites are currently the most investigated class of semiconductors for photovoltaic solar cells.[45-46] When immersed in ethanol, the perovskite film discoloured after few seconds, while it remained stable in the perfluorinated solvent tetradecafluorohexane. It has already been shown that the surface of perovskites are prone to interact with iodo-perfluorocarbons via halogen bonding, [47] suggesting that our approach with IPFC's could also be used to functionalize a perovskite surface. Indeed, it is shown in Figure 4b and c, shows that the equilibrium contact angle of isopropanol on a perovskite substrate changes from 0 to 36° when the its surface is functionalized with I-PFC12. As discussed in Figure 3 for $SiN_x$, such a large change in contact angle indicates the presence of a layer of I-PFC12 on the surface of perovskite. Functionalizing the surface of perovskites with long perfluorocarbon chains is particularly useful to make the material more stable when exposured to humid air, which is currently one of the main limitations for their application in solar cells. Furthermore, it is expected that the electron withdrawing character of the halogen-bond interaction can be used to affect the electron Fermi level of perovskite at the interface with other semiconductors. This and other potential applications of halogen-bond driven self-assembled monolayers in optoelectronics are currently under investigation in our laboratory. We expect that the halogen bonding can be used to functionalise any organic-inorganic surface, as long as electron-donor atoms are available at the exposed surface.

In conclusion, this study has demonstrated a new mechanism for the creation of perfluorinated self-assembled monolayers based on non-covalent halogen bonding as a molecule-to-substrate interaction. Perfluorinated iodo compounds can adsorb onto silicon nitride substrates through the specific interaction between the terminal iodine and the electron lone pair of substrate nitrogen atoms. Similar halogen interactions have been widely explored between iodine and nitrogen-containing organic molecules to form supramolecular adducts, but this is the first report of molecule-to-substrate interactions yielding stable self-assembled monolayers. The experimental and theoretical evidence presented here shows that I-PFC12 self-assembles on silicon nitride via halogen bonding. I-PFC12 SAMs form uniform and densely packed perfluorocarbon coatings, which result in the lowest reported energy for a solid surface of 2.6 mJ $m^{-2}$. Given the low strength of the halogen bond relative to covalent interactions, it is remarkable that these monolayers are stable in fluorinated and aprotic solvents. I-PFC12-based SAMs might therefore be useful to modify a wide range of inorganic and organic substrates and particles, as long as electron-donor atoms are present on the solid surface. We expect these halogen bond stabilized monolayers will find applications in organic thin film electronics, such as perovskite solar cells, where the use of aprotic and fluorinated solvents is of interest to improve the stability of the organic semiconductors.[48]

**Experimental Section**

*Modelling:* The absorption of perfluorododecyl iodide (I-PFC12) on the (001) surface of $SiN_x$ was studied by using density functional theory (DFT).[49] The calculations were performed using the generalized gradient approximation (GGA) of Perdew-Burke-Ernzerhof (PBE)[50] for the exchange and correlation functional, plus van der Waals interaction corrections[51] in order to account for the non-covalent interaction between the I-PCF12 molecules and the surfaces. The electron-ion interactions were computed using ultrasoft pseudopotentials, including scalar relativistic effects. The energy cut-off for the plane wave expansion was 30 Ry (300 Ry for the charge density cutoff). Brillouin-zone integration was performed with the special-point technique,[52] which was done with a 6 × 6 × 1 Monkhorst-Pack grid. The (001) semiconductor surfaces were modelled by employing a periodically repeated slab geometry using the (1 × 1) supercell for silicon nitride, with adsorbates on one side of the slab only. The slabs were 5 layers thick and the bottom layers were kept fixed in their bulk positions during relaxations. A 12 Å vacuum layer was used, which was found to be sufficient to ensure negligible coupling between periodic replicas of the slab. All calculations were performed using the PWscf code contained in the Quantum-ESPRESSO distribution.[53]

*Surface preparation.* Silicon substrates coated with about 300 nm of silicon oxide and 50 nm silicon nitride (Pi-Kem Ltd, UK) were cleaned by sonication for several minutes in distilled ethanol and acetone. After drying the surface in a high purity nitrogen flow, they were further cleaned by a $CO_2$ snow jet (Applied Surface Technologies, NJ, USA). Then, an air plasma treatment (Harrick Plasma Cleaner, Ithaca NY, model PDC-002) was applied for 1 min at 30 W. Freshly cleaned substrates were immersed in a 10 mM solution of perfluorododecyl iodide (I-PFC12, Sigma Aldrich) in tetradecafluorohexane (Fluorinert® FC-72, 3M, USA) for about 30 minutes. Immediately after removing the samples from the I-PFC12 solution, they were rinsed with tetradecafluorohexane and then immersed for several hours in a large volume of fresh tetradecafluorohexane. After drying the surface in a high purity nitrogen stream the samples were ready for characterisation.
Perovskite substrates were prepared according to previous reports.[45] Freshly prepared perovskite films were treated with I-PFC12 following the same method described for silicon nitride substrates.

*Characterisation.* Contact angle measurements (KSV Instruments Ltd CAM 200) were taken using 5 µl droplets of different alkanes; for advancing and receding contact angles, the dispensing rate was 0.2 µl/s. High purity alkanes (pentane, hexane, heptane, decane, dodecane and hexadecane) were used as received from Aldrich or Merck. Tapping Mode AFM (Bruker, Multimode 8) was used to image surface topography.

*X-ray reflectivity (XRR)* data were measured at beamline D1, at the Cornell High Energy Synchrotron Source (CHESS) at Cornell University, USA. The wavelength was $\lambda$ = 1.15 Å. XR measurements were performed *in situ* using collimating slits and the goniometer and sample environment of the grazing incidence small-angle X-ray scattering (GISAXS) geometry. The detector was an ion chamber with a 50 × 13 $mm^2$ aperture. A blade placed in front of the ion chamber screened the detector from the direct beam at low angles. The measuring time was 1 s per data point. The electronic background was measured and subtracted from the data.

*X-ray photoelectron spectroscopy (XPS)* data collection was performed in ultra-high vacuum ($<10^{-9}$ mbar and analyses were performed with an Escalab 220i-XL spectrometer (VG, UK) with monochromatic Al K$\alpha$ exciting radiation (energy 1486.6 eV). Typical operating conditions were: power: 120 W (10 kV, 12 mA) and spot size: 1 $mm^2$. The survey spectra of Figure 2a were collected with a step width of 1 eV and high-resolution spectra of Figure 2 b,c with a step width of 0.1 eV.

*Silicon nitride nanoparticles* for $^{19}F$ ssNMR (UBE Industries, LTD, SNE-10, nominal range 0.5 µm) were utilized in $^{19}F$ solid-state NMR spectroscopy (ssNMR) to model the adsorption of IPFC-12 monolayers through halogen-bonding. SAM-covered nanoparticles were prepared by immersing them in 10 mM solution of perfluorododecyl iodide (IPFC12) in tetradecafluorohexane for 20 hours in a rotary mixer (Ratek, BTR10-12V). The nanoparticles were subsequently centrifuged at 6000 rpm for 10 min. After removal of the solvent,The supernatant was removed, and the nanoparticles were left to dry for 10 minutes at ambient temperature in air. Fresh tetradecafluorohexane. Tetradecafluorohexane was used to rinse the nanoparticles, to remove any unbound IPFC-12, then recoveredfollowed by centrifugation at 6000 rpm for 10 minutes, followed by and drying in a stream of nitrogen for 10 minutes. The washing/centrifugation/drying cycle was repeated four more5 times to remove residual unbound I-PFC12.

*$^{19}F$ ssNMR* data were acquired with a Bruker Avance III 700 MHz spectrometer operating at 658.9 MHz. All samples were packed into Bruker 1.3 mm outer diameter $ZrO_2$ rotors and set to spin at 60 kHz in a 1.3 mm MAS probe allowing a sample MAS of 60 kHz. Trifluoroacetamide (Sigma Aldrich) was used to tune the NMR probe to $^{19}F$ frequency, and as an external reference ( - , .

**Supporting Information**

Supporting Information available from the Wiley Online Library.

**Acknowledgements**

A.A., A.S. and U.S. acknowledge support from the Adolphe Merkle Foundation. C.N. acknowledges funding from the University of Sydney and the Australian Research Council. AA has received funding from the European Union's Seventh Framework Programme for research, technological development and demonstration under grant agreement no 291771. RD acknowledges the EPSRC EP/G060649/1for funding. Part of the work was conducted at beam line D1 at the Cornell High Energy Synchrotron Source (CHESS), CHESS is supported by the NSF & NIH/NIGMS via NSF award DMR-1332208. We thank D. Smilgies, X. Sheng and J. Dolan for their help during the D1 experiment at CHESS. The NMR spectrometers in the Mark Wainwright Analytical Centre, UNSW, were supported by Australian ARC LIEF grants (LE0989541 for solids & LE120100027 for solutions), and access is gratefully acknowledged.

# Supporting Information

# Halogen-Bond Driven Self-Assembly of Perfluorocarbon Monolayers

# Contents

**Modelling the monolayer on a silicon substrate**

The absorption of perfluorododecyl iodide (I-PFC12) on the (001) surface of $SiN_x$ was studied by using density functional theory (P. Hohenberg and W. Kohn; Phys. Rev. 136, B864; 1964). The calculations were performed using the generalized gradient approximation of Perdew-Burke-Ernzerhof (John P. Perdew, Kieron Burke, and Matthias Ernzerhof; Phys. Rev. Lett. 77, 3865, 1996) for the exchange and correlation functional, plus van der Waals interaction corrections (R. Sabatini, T. Gorni, and S. de Gironcoli Phys. Rev. B 87, 041108(R), 2013) in order to account for the non-covalent interaction between the I-PFC12 molecules and the surfaces. The electron-ion interactions were computed using ultrasoft pseudopotentials, including scalar relativistic effects. The energy cut-off for the plane wave expansion was 30 Ry (300 Ry for the charge density cutoff). Brillouin-zone integration was performed with the special-point technique (H. J. Monkhorst and J. D. Pack Phys. Rev. B 13, 5188, 1976) which was done with a 6 × 6 × 1 Monkhorst-Pack grid. The (001) semiconductor surfaces were modelled by employing a periodically repeated slab geometry using the (1 × 1) supercell for silicon nitride, with adsorbates on one side of the slab only. The slabs were five layers thick, and the bottom layers were kept fixed in their bulk positions during relaxations. A 12 Å vacuum layer was used, which was found to be sufficient to ensure negligible coupling between periodic replicas of the slab. All calculations were performed using the PWscf code contained in the Quantum-ESPRESSO distribution (J. Phys.: Condens. Matter 21 (2009) 395502).

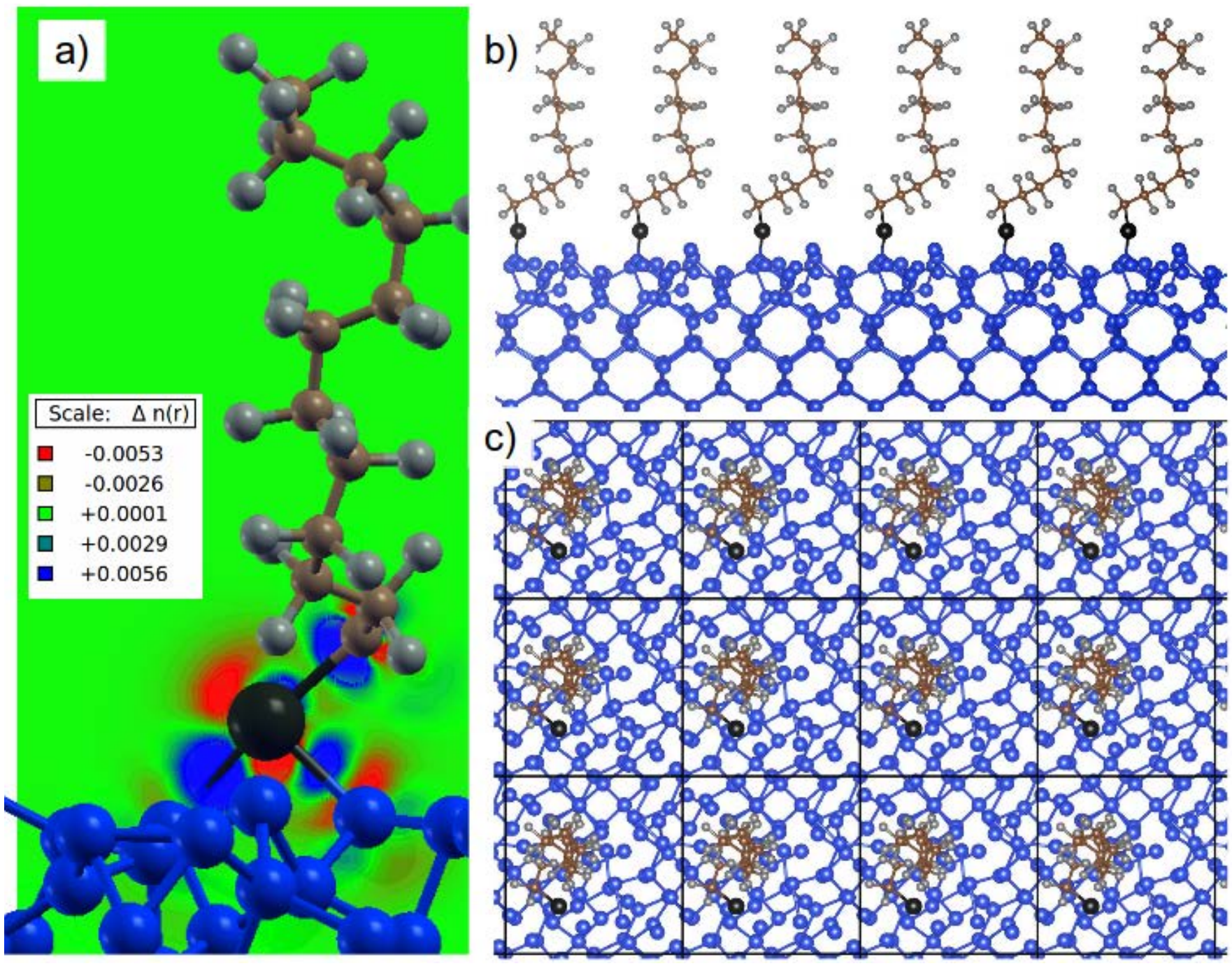

a) Contour plot of the charge density defined as $\Delta n(r) = n_{tot} - [n_{IPFA} + n_{surf}]$ of perfluorododecyl iodide (I-PFC12) on a Si (001) surface, linear scale from −0.005 to 0.005 e/Å$^3$ with the increment of 0.003 e/Å$^3$. The charge flows from the red to blue region. The analysis shows that there is significant charge density reorganisation induced by hybridisation between the valence electrons of the iodine and silicon atoms, which indicates the formation of a covalent interaction between IPFA and the surface. (b) Side view and (c) top view of a self-assembled monolayer of I-PFC 12 onto Si (001) surface as calculated from DFT method. Elements colour code: I, black; C, brown; F, white; Si, blue.

## Halogen bond on $SiO_2$ substrate

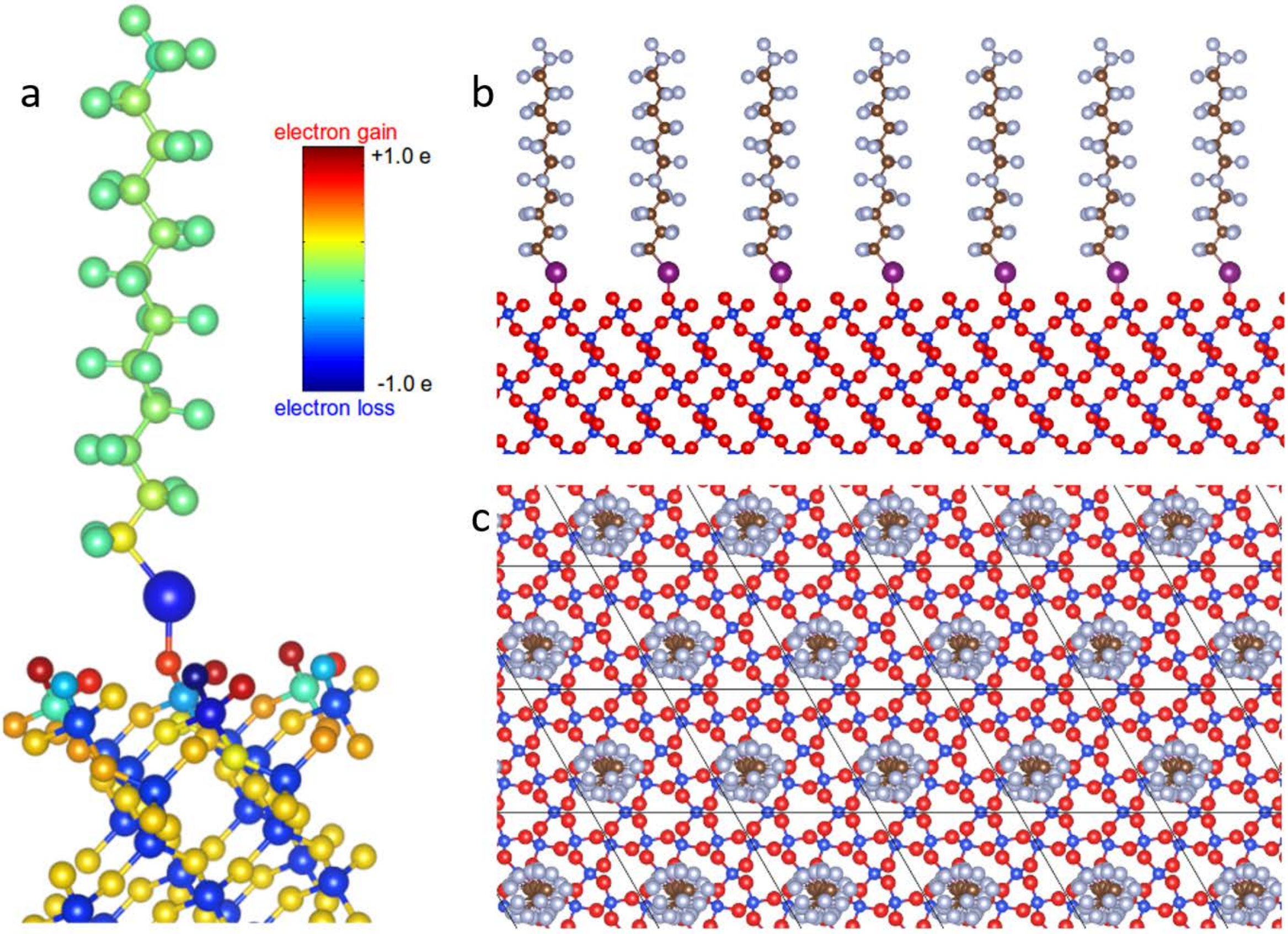

a) Löwdin atomic charge analysis of perfluorododecyl iodide (I-PFC12) on an alpha-SiO2(001) surface as calculated from density functional theory (see Experimental Section for more details). The terminal I atom the I-PFC12 is linked to the O atom on the substrate surface. The colour of the atoms indicates the electron density redistribution upon the interaction of I-PFC12 with the silica substrate: red (blue) corresponds to electron gain (loss). b) Side view and (c) top view of I-PFC12 on a silicon oxide substrate with the I-PFC12 molecules on the surface as calculated by density functional theory. Element colour codes in b and c: I, violet; C, brown; F, white; O, red; Si, blue.

From the Figure a, we see a significant charge transfer between of the on-surface oxygen atoms and I-PFC12 molecule. However, different from the case of halogen-bond between I-PFC12 and $SiN_x$ surface, where the electron transfer takes place from N to I, for the case of I-PFC12 on $SiO_2$, the electron density moves from I to O atoms on the surfaces. This result is consistent with the so-called multivalent halogen-oxygen bonds, where the lower electronegativity of I compared to O allows iodine to carry a much larger positive charge distribution and enhances the ionic nature of I–O bonding (Bioinorg Chem Appl. 2007; 2007: 46393).

**Modelling the trans conformation of I-PFC12 on SiNx**

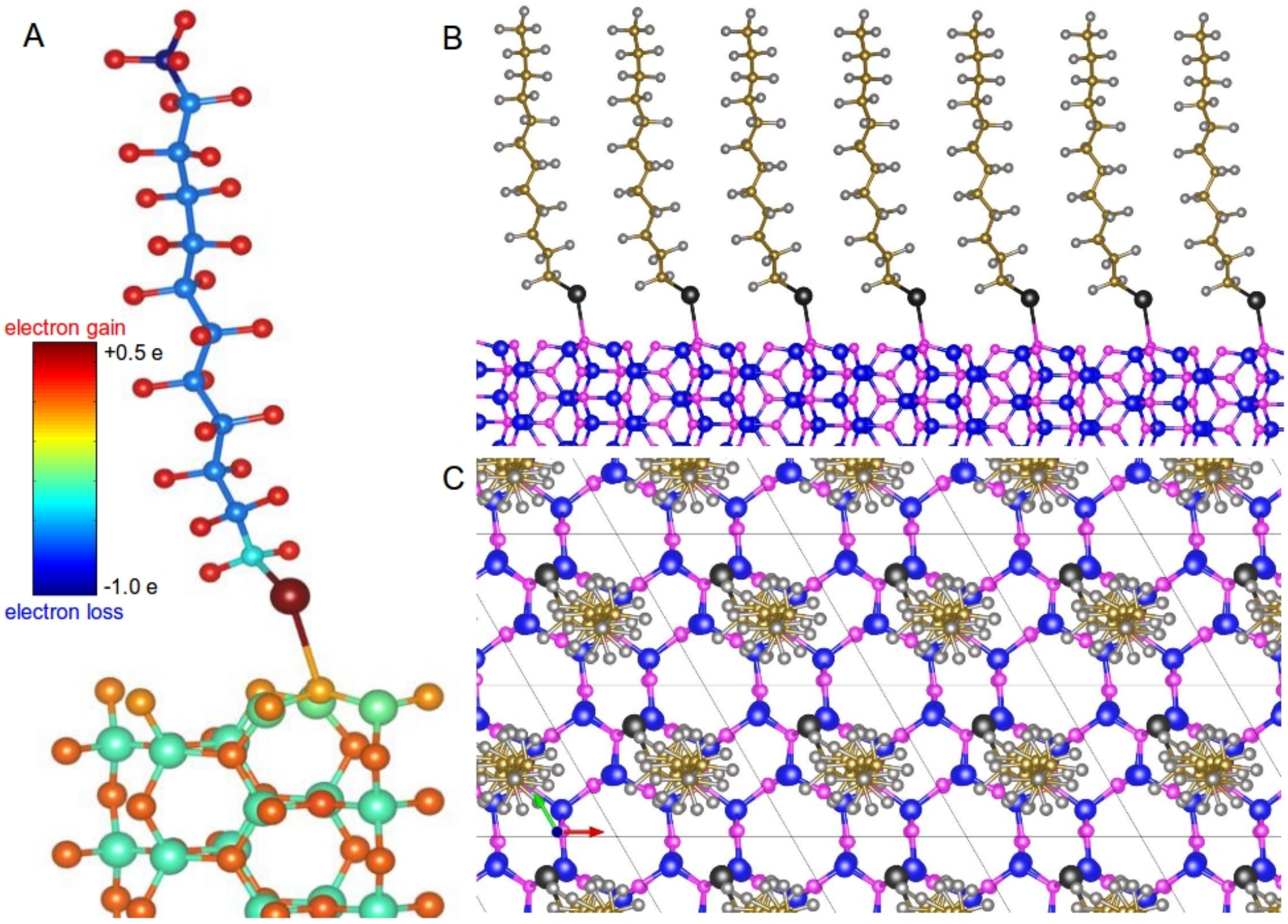

a) Löwdin atomic charge analysis of perfluorododecyl iodide (I-PFC12) on a (001) silicon nitride surface as calculated from density functional theory (see Experimental Section for more details). The terminal I atom the I-PFC12 is linked to the N atom on the substrate surface. The colour of the atoms indicates the electron density redistribution upon the interaction of I-PFC12 with the silicon nitride substrate: red (blue) corresponds to electron gain (loss). b) Side view and (c) top view of I-PFC12 on a silicon nitride substrate with the I-PFC12 molecules on the surface as calculated by density functional theory. Element colour codes in b and c: I, black; C, brown; F, white; N, violet; Si, blue.

We calculated I-PFC12 on $SiN_x$ surface with a fully trans conformation. From the calculations, we found that with the new conformer:

The height of the monolayer from the model of I-PFC12 on $SiN_x$ (001): 17.12 Å

Averaged C-C bond: 1.623 Å

Averaged C-F bond: 1.307 Å

C-I bond length: 2.190 Å

I-N bond length: 2.805 Å

The binding energy: I-PFC12-$SiN_x$ (001) surface: 0.92 eV (88 kJ/mol).

The charge distribution does not change very much compared to the gauche conformation reported in the main text (see Figure 1a). However, the binding energy (88 KJ/mol) is lower, thus suggesting the gauche conformation is more favourable.

## Atomic force microscopy

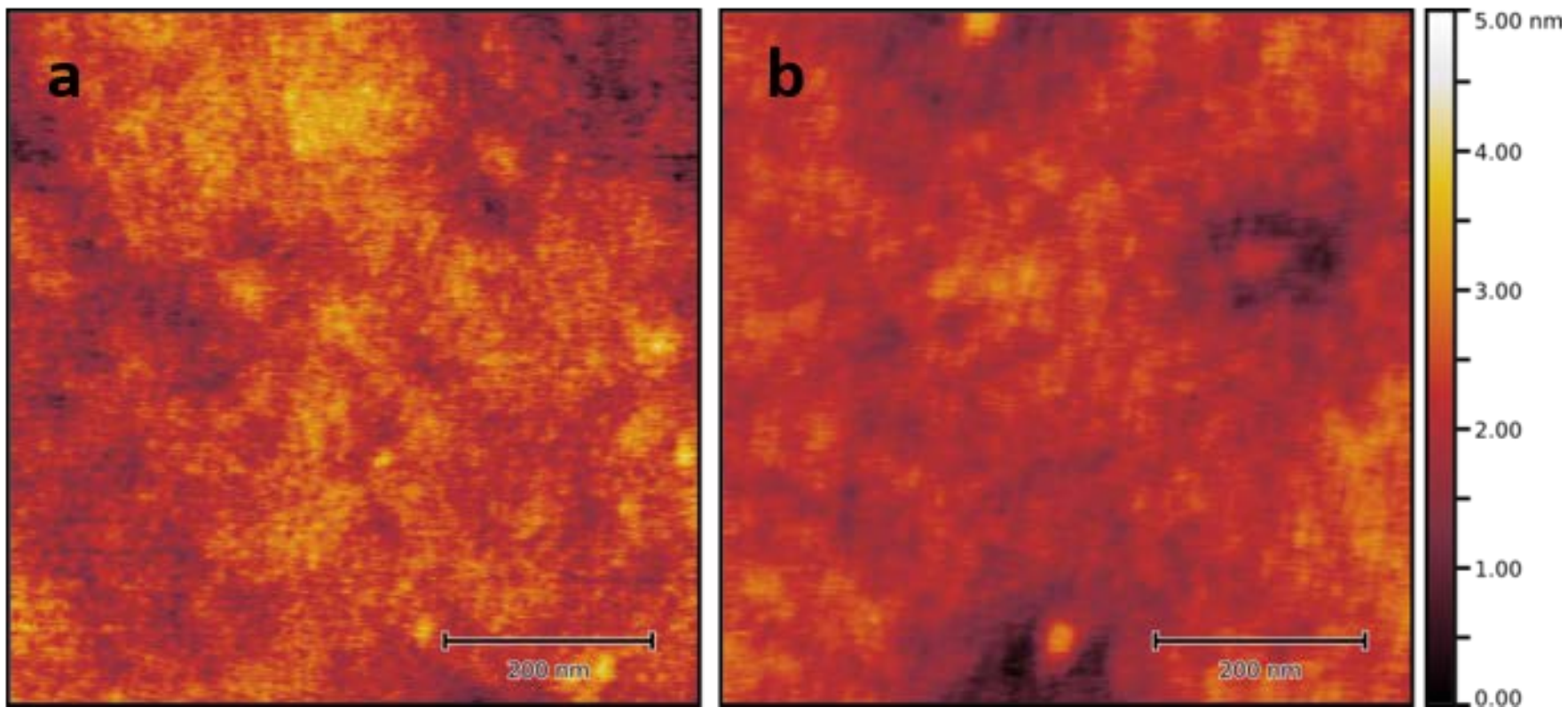

Tapping Mode™ AFM topography images of $SiN_x$ substrates. a) The plain silicon nitride substrate was cleaned by sonication for several minutes in distilled ethanol after a thorough cleaning to remove particulate and acetone. After drying the surface with a high purity nitrogen flow, it was further cleaned by a $CO_2$ snow jet, and then an air plasma treatment was applied. b) After this cleaning procedure it was immersed in the I-PFC12 solution in tetradecafluorohexane, and then rinsed for several hours in a large volume of fresh tetradecafluorohexane. The height scale applies to both images. Both the samples show a mean rms surface roughness of ca. 0.5 nm. This suggests that either no molecules have adsorbed onto the substrate, or that adsorbed molecules have formed a uniform and compact layer, conformably attached to the substrate surface.

**Elemental composition extracted from the XPS spectra in Figure 2 of the main text**

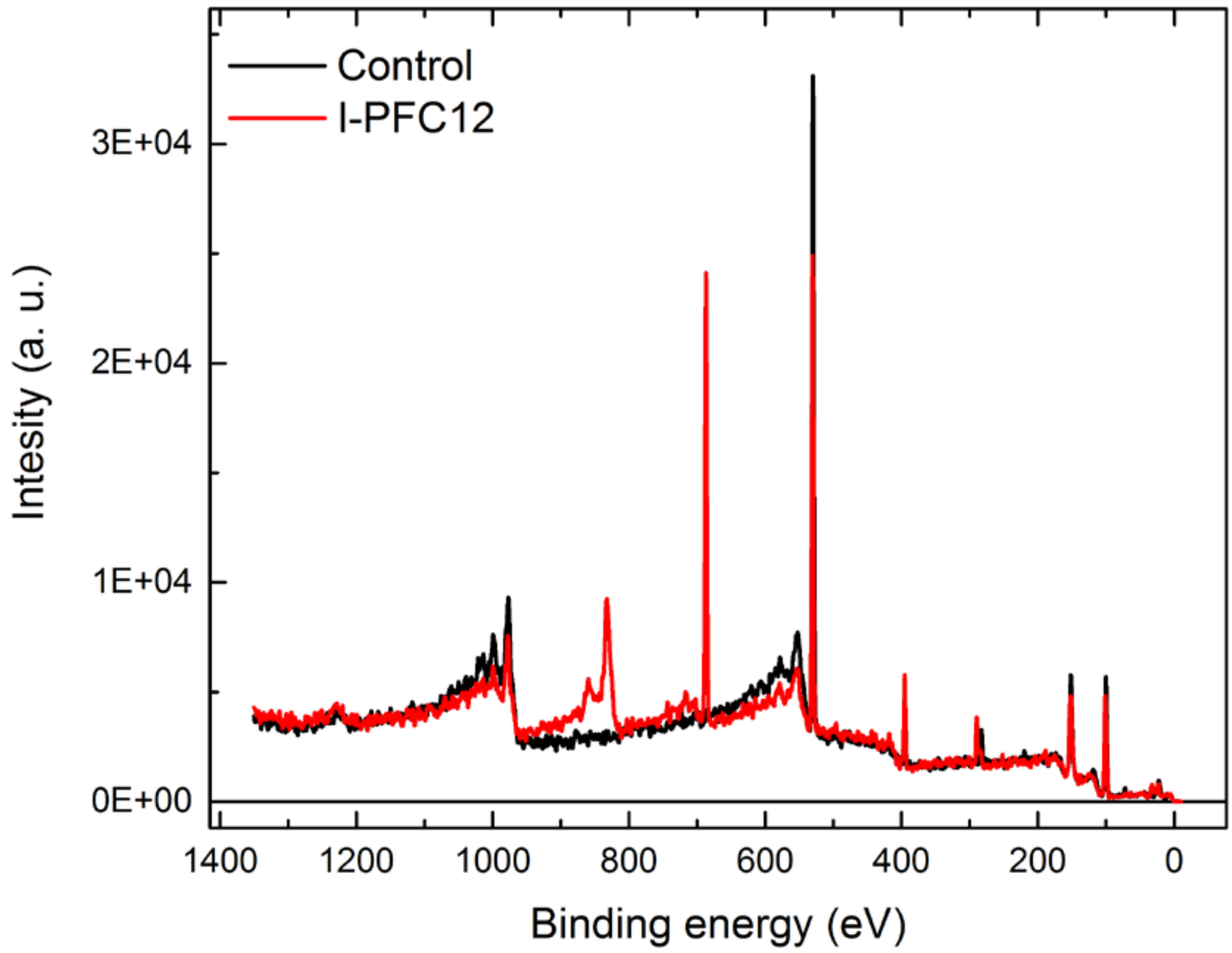

| Silicon nitride immersed in I-PFC12 | | | |
|---|---|---|---|
| Label | Peak binding energy (eV) | Atomic content (%) | Total atomic content (%) |
| O1s A | 530.38 | 7.43 | 31.87 |
| O1s B | 530.99 | 13.09 | |
| O1s C | 531.64 | 11.35 | |
| Si2p A | 100.21 | 10.92 | 26.08 |
| Si2p B | 101.65 | 15.16 | |
| C1s A | 283.99 | 2.6 | 12.02 |
| C1s B | 287.56 | 1.39 | |
| C 1s C | 289.92 | 7.32 | |
| C1s D | 292.25 | 0.71 | |
| N1s | 396.01 | 9.3 | 9.3 |
| F 1s | 687.27 | 20.73 | 20.73 |

**Grazing-Angle FTIR**

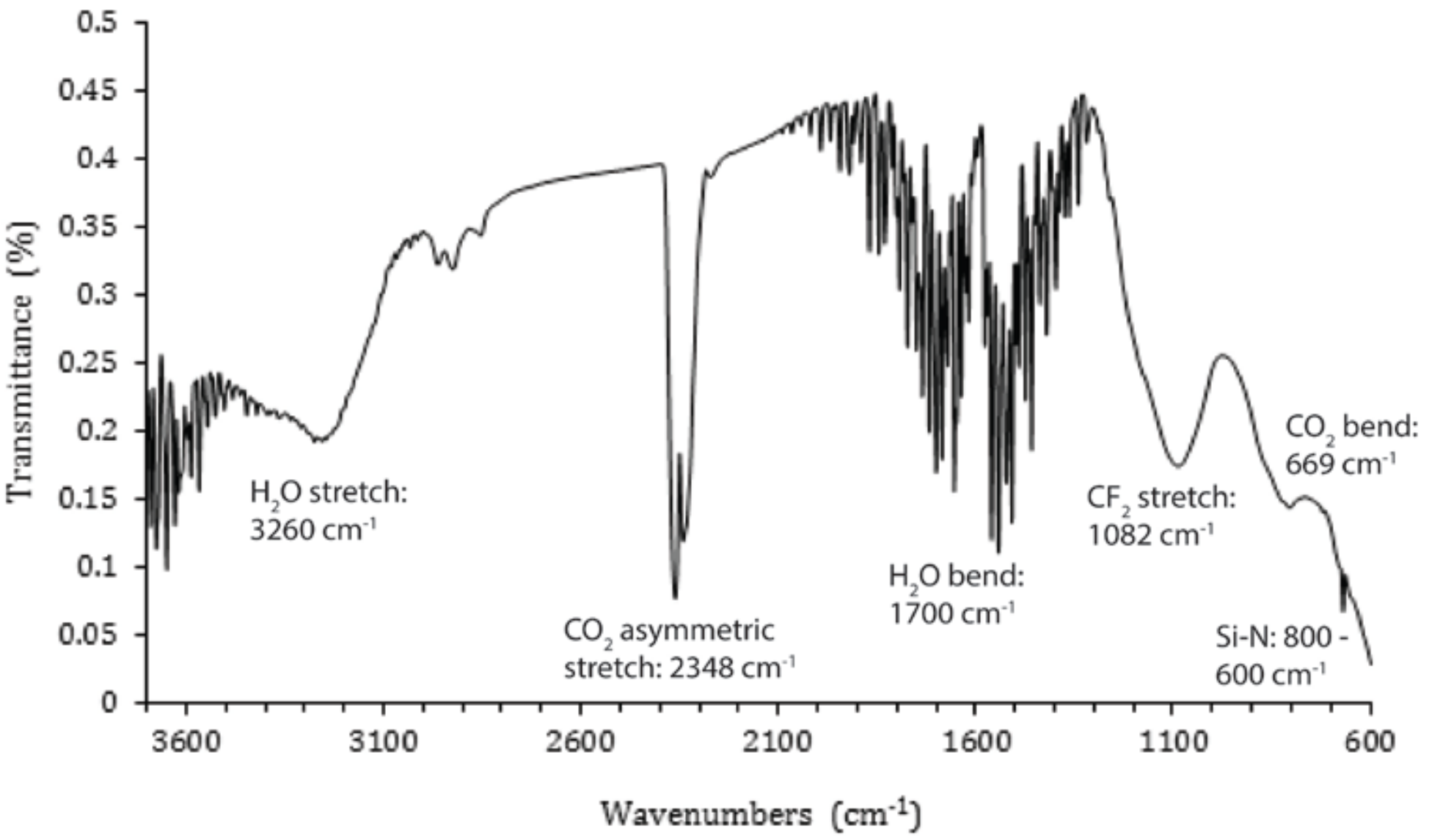

Specular reflectance grazing-angle FTIR spectrum of I-PFC12 halogen-bonded to silicon nitride with at a grazing-angle of 84°. Resolution of 4 cm$^{-1}$ with 512 scans.

The large noise present in the spectrum is likely due to optical effects caused by the low reflectance of silicon at grazing angles near its Brewster's angle, and this prevents the collection of detailed information on the monolayer (Appl. Spectrosc. 2007, 61(5), 530-536). Evidence of the presence of a fluorinated layer is seen in the C-F stretching found at 1082 cm$^{-1}$ (literature values 1200 – 1000 cm$^{-1}$, Appl. Spectrosc. 2007, 61(5), 530-536); the C-I signal is probably hidden by the silicon nitride peaks (800 – 600 cm$^{-1}$, Appl. Spectrosc. 2007, 61(5), 530-536). Adventitious water and carbon dioxide were also observed ($H_2O$ stretching at 3260 cm$^{-1}$ and bending at 1700 cm$^{-1}$; carbon dioxide asymmetric stretch at 2348 cm$^{-1}$ and degenerate bend at 669 cm$^{-1}$, Appl. Spectrosc. 2007, 61(5), 530-536; Aust. J. Phys., 1982, 35, 623-38).

Grazing-Angle FTIR was carried out using a Bruker Tensor 27 instrument with a KBr beamsplitter and MCT with a Hyperion 3000 FT-IR microscope and Grazing-Angle Objective (Bruker Optics, Inc.) at a grazing-angle of ~84°.

**$^{19}$F CP-MAS NMR spectrum of trifluoroacetamide as an internal standard**

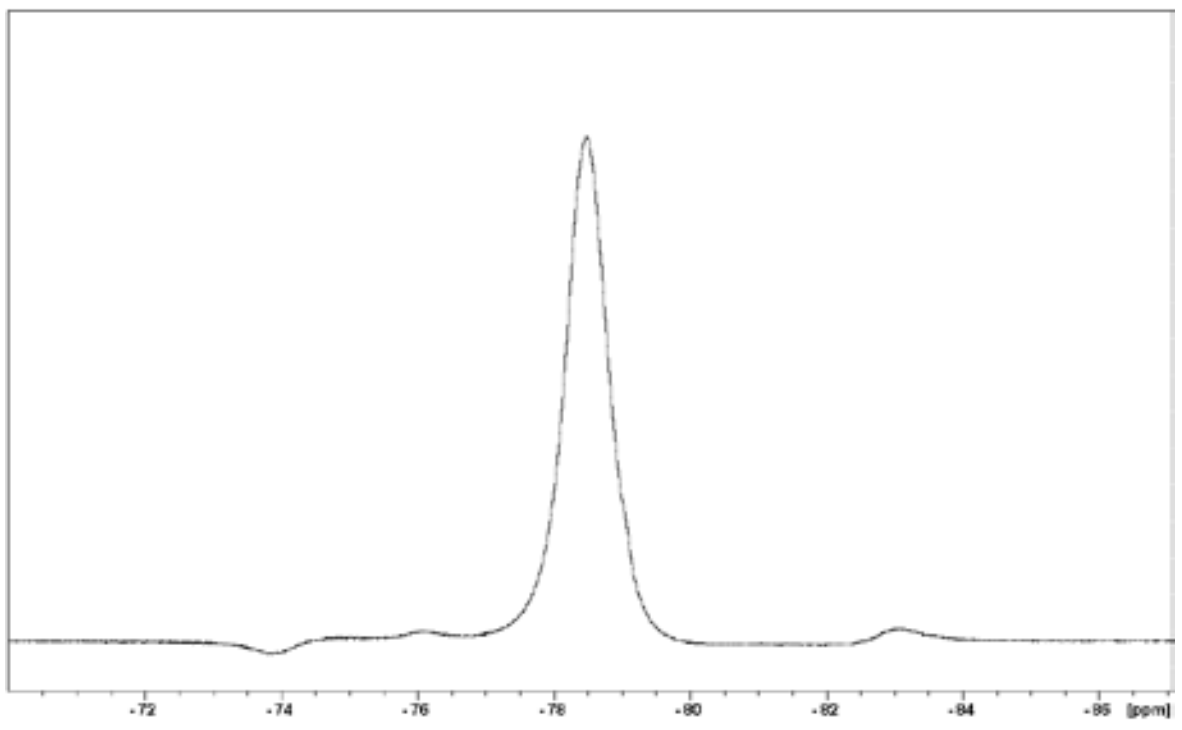

$^{19}$F CP-MAS NMR spectrum of trifluoroacetamide as internal standard taken with the Bruker Avance III 700 MHz spectrometer at 658.9 MHz (60 kHz MAS) Chemical shift of Trifluoroacetamide, δ = -78.49 ppm, used as a standard for NMR probe tuning to the $^{19}$F isotope frequency in a 1.3 mm CPMAS probe.

**XPS spectra without plasma treatment**

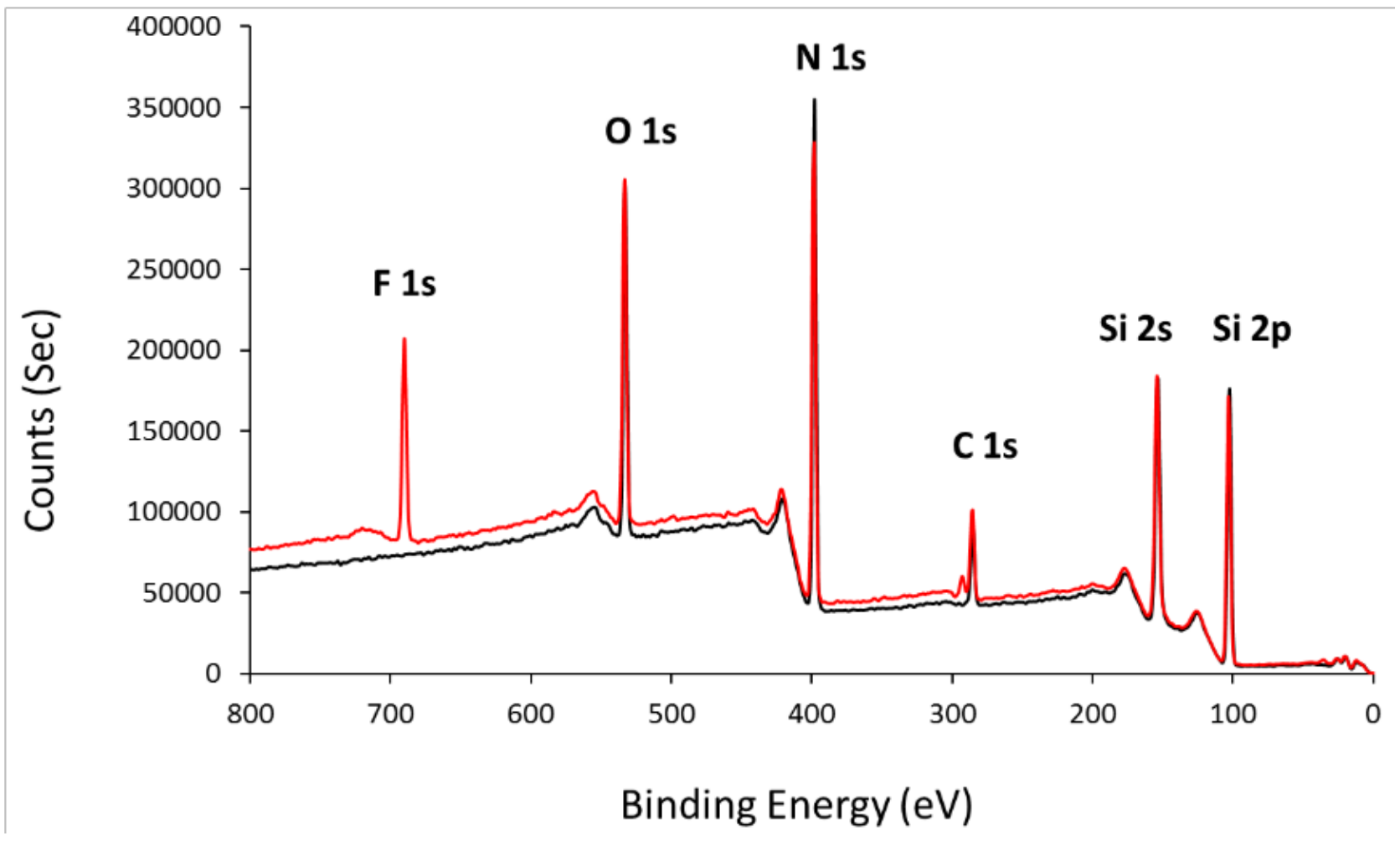

XPS spectra of silicon nitride substrates that were immersed in a solution of tetradecafluorohexane with and without (black) perfluorododecyl iodide (red). Different from the spectra in the main text, the nitride substrates were not treated with the plasma cleaning before the functionalization.

| Silicon nitride immersed in I-PFC12 (without plasma treatment) | | | |
|---|---|---|---|
| Label | Peak binding energy (eV) | Atomic content (%) | Total atomic content (%) |
| O1s | 532.52 | 15.02 | 15.02 |
| Si2p A | 102.02 | 27.95 | 33.61 |
| Si2p B | 103.13 | 5.66 | |
| C1s A | 284.8 | 9.54 | 12.94 |
| C1s B | 286.48 | 0.94 | |
| C 1s C | 287.8 | 0.15 | |
| C1s D | 289.46 | 0.3 | |
| C1s E | 292.04 | 1.8 | |
| C1s F | 294.39 | 0.21 | |
| N1s | 397.8 | 32.6 | 32.6 |
| F 1s | 689.33 | 5.85 | 5.85 |

**XPS depth profile of silicon nitride substrates non-plasma treated**

The XPS depth profile for the nitride substrate, revealing around 5% oxygen content in the top layer. Therefore, we cannot exclude that oxygen is also involved in the interaction of the silicon nitride surface with the iodine of the I-PFC12. Nevertheless, we note that the nitrogen content in the top layer is significantly higher (40%), suggesting that oxygen is not forming a uniform top layer that may prevent from a direct interaction between the I-PFC12 and the nitrogen.

Analytical Technique: X-Ray Photoelectron Spectroscopy (XPS)
Instrument: ESCALAB250Xi
Manufacturer: Thermo Scientific, UK
Background vacuum: better than 2E-9 mbar
X-ray source: mono-chromated Al K alpha (energy 1486.68 eV)
Power: 150W (13 kV x 12 mA)
Spot size: 500 micrometres
Photoelectron take-off angle: 90 degrees
Pass energy: 100 eV for survey scans, or 50 eV for depth profiling region scans
Software: Avantage
Spectrometre calibration: Au 4f7 = 83.96 eV, Ag 3d5 = 368.21 eV, Cu2p3 = 932.62 eV

## $^{19}$F solution-state NMR

We investigated the potential covalent attachment of the I-PFC12 on silicon nitride substrates performing the following experiments:

1. I-PFC12 molecules were adsorbed onto silicon nitride microparticles and then washed off with deuterated acetone. The nature of the recovered molecules was unchanged, as shown from the $^{19}$F solution-state NMR reported below:

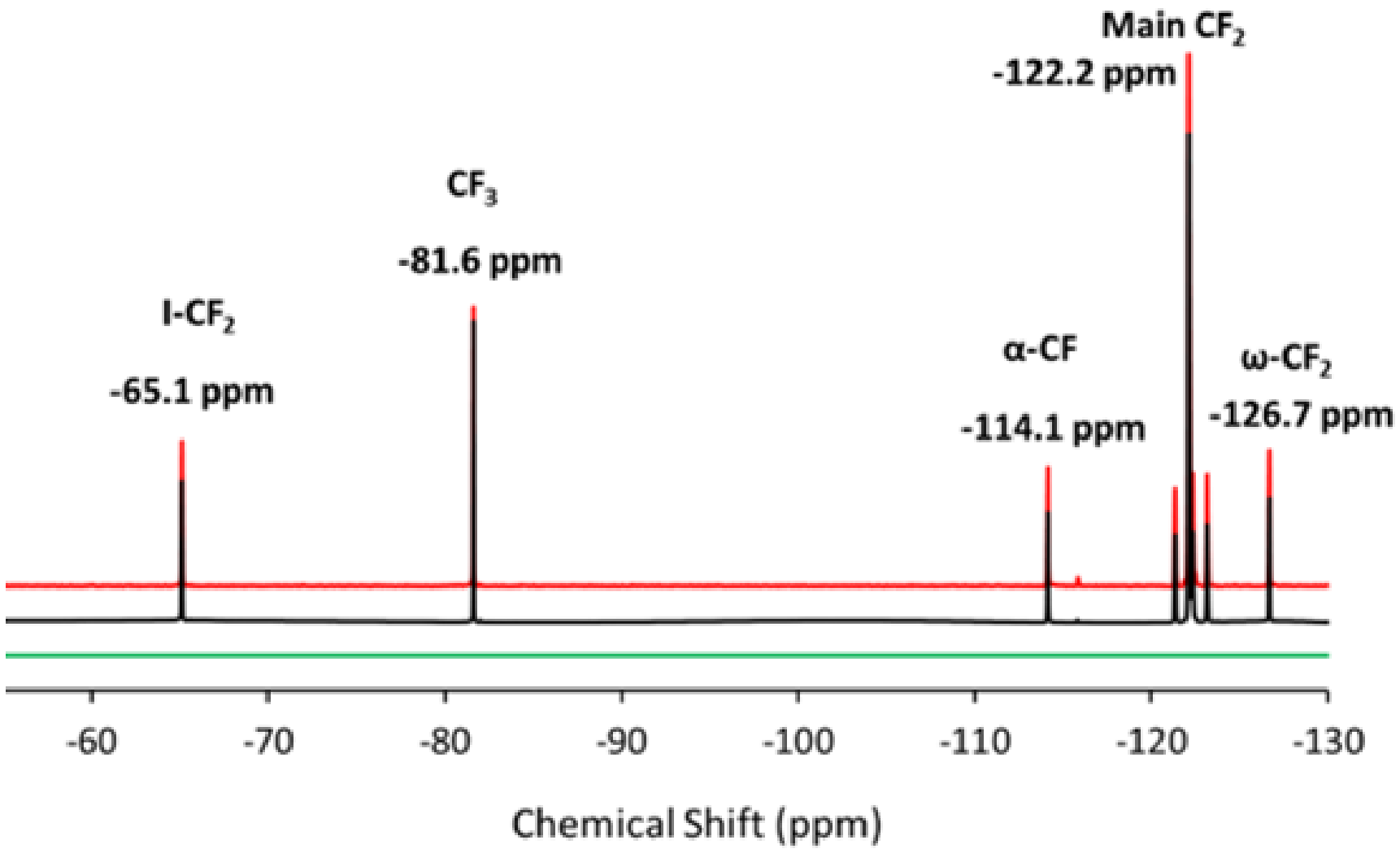

$^{19}$F solution-state NMR spectra of I-PFC-12 dissolved in acetone-$d_6$ with baseline correction (black) after removal from $Si_3N_4$ nanoparticles; of the pure I-PFC12 solution in (red); solution removed from pure $Si_3N_4$ nanoparticle non-functionalised but washed with pure solvent only, perfluorohexane (green). Measurements taken on 600 MHz spectrometer with Pre-scan delay = 18 µs and D1 = 8 µs over 128 scans.

$^{19}$F NMR chemical shift assignments of IPFC-12 in acetone-d6 as a pure solution and recovered from silicon nitride nanoparticles; measured on 600 MHz spectrometer. Literature values for $ICF_2$– on 1,8-diiodoperfluorooctane, halogen bonded to acetone in solution quoted in brackets (J. Fluorine Chem. 2002, 114, 27–33).

| I-$CF_2$ (ppm) | $CF_3$ (ppm) | α-$CF_2$ (ppm) | Main $CF_2$ (ppm) | ω-$CF_2$ (ppm) |
|---|---|---|---|---|
| 65.1 (65.5) | 81.6 | 114.1 | 122.2 | 126.7 |

2. I-PFC12 molecules were dissolved off a silicon nitride wafer after the functionalized wafer had been stored on a laboratory bench in ambient light for three days. The molecules dissolved off the wafers did not resemble the starting I-PFC12, but rather a similar compound without the iodide attached. This was deduced from the presence of peaks consistent with $CF_3(CF_2)n$, but missing the peaks at -65 and -114 ppm, indicative of the I-$CF_2$-$CF_2$- group (see figure a reported below).

In contrast, when the silicon nitride wafer was stored in the dark before NMR measurements, the recovered molecule was the unchanged I-PFC12 (see figure b reported below).

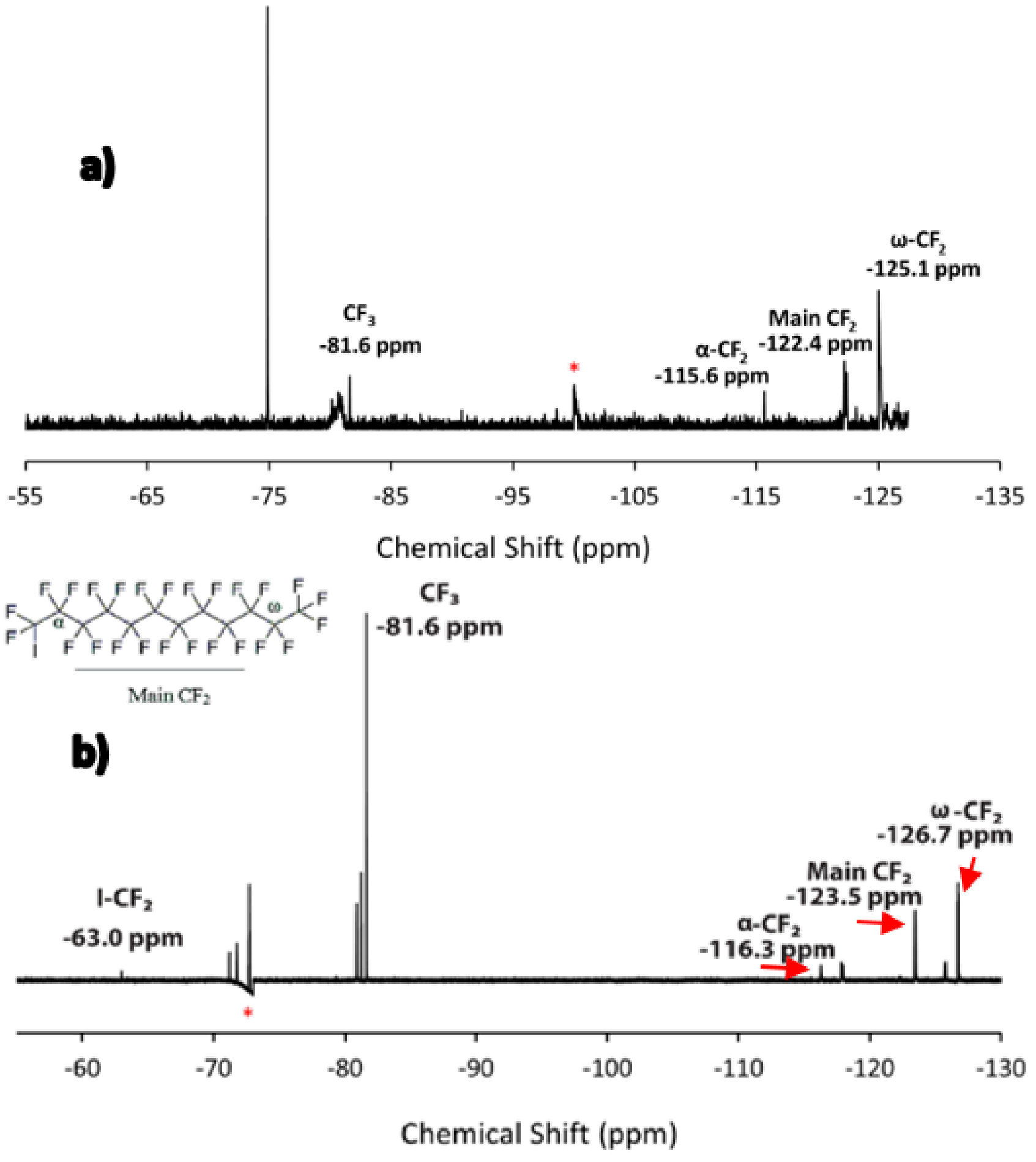

a) $^{19}F$ solution-state NMR spectra of I-PFC12 dissolved in acetone-$d_6$ as upon removal from a silicon nitride wafer stored on a lab bench, exposed to light for three days. Baseline correction artefacts indicated with asterisks. Measurements taken on 600 MHz spectrometer with pre-scan delay = 18 µs, D1 = 8 µs over 128 scans. b) $^{19}F$ solution-state NMR spectra of I-PFC12 dissolved in acetone-$d_6$ upon removal from a silicon nitride wafer stored in the dark. Baseline correction artefacts indicated with an asterisk. Measurements taken on 600 MHz spectrometer with Pre-scan delay = 18 µs, D1 = 8 µs over 1024 scans.

The results of the NMR analysis suggest that photo-induced dissociation of the I-PFC12 molecule occurs on the surface of the silicon nitride wafers upon exposure to light. The differences between the signals found for the wafers and nanoparticles is likely because the majority of the nanoparticles are buried away from the light while the surface of the wafers is entirely exposed to ambient light. As can be seen in the figure reported below, I-PFC12 molecules in a perfluorohexane solution are unchanged over one month of exposure to light, suggesting substrate-mediated photolysis as described in other studies. Other examples of the photo-dissociated detachment of iodine exist, such as the finding that photoemission of electrons from the substrate upon exposure to UV light induces the cleavage of the C-I bonds of iodo-perfluoromethane on surfaces of silver (Wang, Y.; Wang, J.; Li, G. X.; He, G.; Chen, G. Halogen-Bond-Promoted Photoactivation of Perfluoroalkyl Iodides: A Photochemical Protocol for Perfluoroalkylation Reactions. Org. Lett. 2017, 19, 1442–1445.)

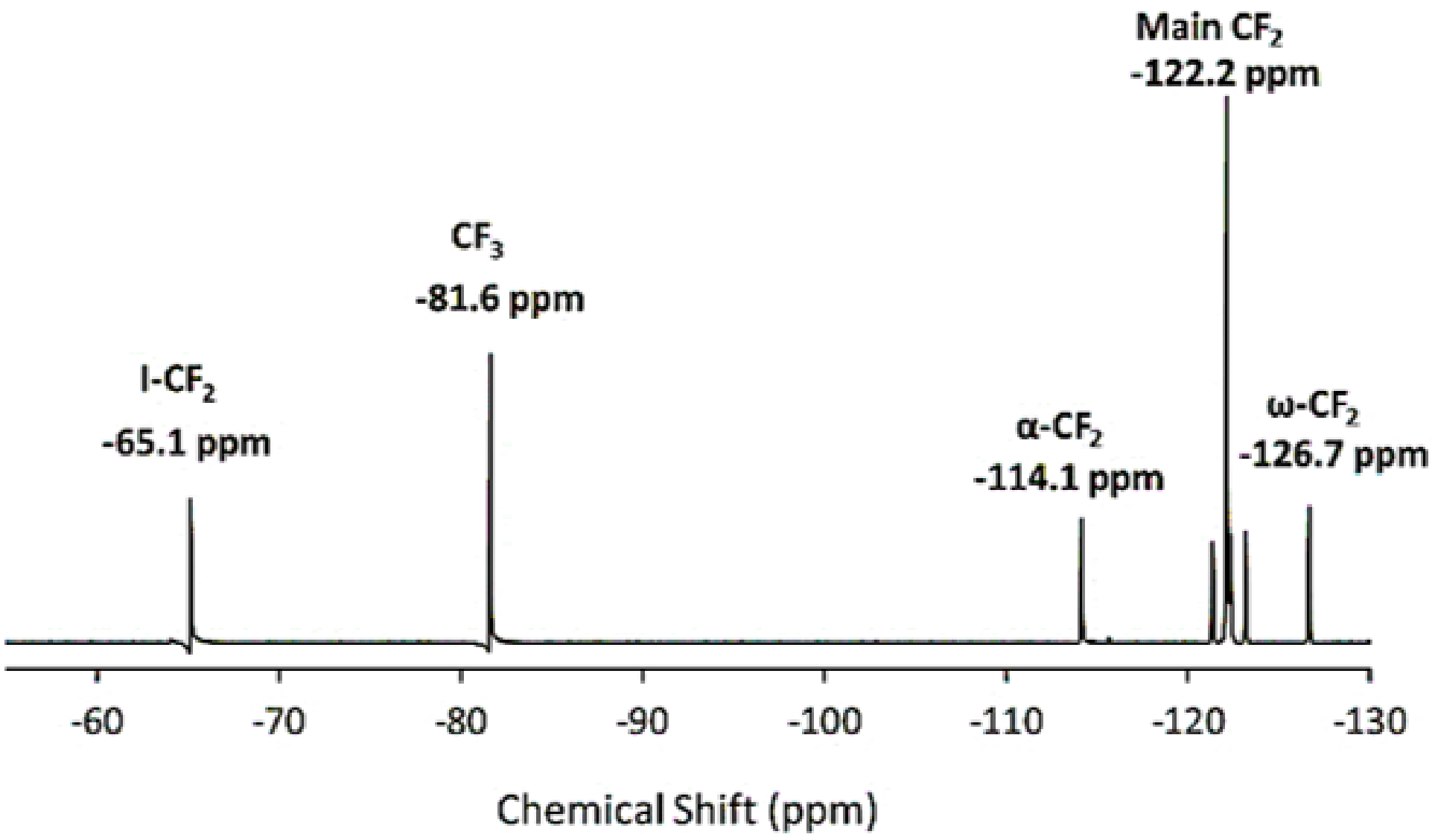

$^{19}F$ solution-state NMR spectra of IPFC-12 dissolved in acetone-$d_6$ with exposure to light for one month with baseline correction. Measurements taken on 600 MHz spectrometer with Pre-scan delay = 18 µs, D1 = 8 µs, over 128 scans.